\documentclass{aa}  

\usepackage{graphicx,subfigure}
\usepackage{booktabs}
\usepackage{ulem}

\usepackage{tablefootnote}

\usepackage{booktabs,caption}
\usepackage[flushleft]{threeparttable}

\usepackage{txfonts}

\usepackage{hyperref}

\begin{document}

   \title{
   Insight into the occurrence of particle acceleration through the investigation of Wolf--Rayet stars using uGMRT observations}

   \author{A. B. Blanco
          \inst{1}, 
          M. De Becker\inst{1},
           A. Saha \inst{2},
           A. Tej \inst{2},
           \and
           P. Benaglia \inst{3}
          }

   \institute{Space Sciences, Technologies and Astrophysics Research (STAR) Institute, University of Liège, Quartier Agora, 19c,
Allée du 6 Août, B5c, 4000 Sart Tilman,
               Belgium \\
              \email{ablanco@uliege.be}
              \and
             Indian Institute of Space Science and Technology, Thiruvananthapuram 695 547, Kerala, India 
              \and
              Instituto Argentino de Radioastronomia (CONICET; CICPBA; UNLP), C.C. No 5, 1894, Villa Elisa, Argentina 
              \\
             }


 
  \abstract
   {Massive stars produce strong stellar winds that consist of continuous outflows of material at speeds of thousands of $\rm km\,s^{-1}$. These winds convey large amounts of kinetic power, especially in the case of Wolf--Rayet (WR) stars. When these winds interact with nearby material, they will likely produce shocks. Among other processes, particle acceleration is expected to occur. This is particularly well established in the case of massive binary systems, where the stellar winds collide, allowing these systems to be identified thanks to the detection of synchrotron radio emission, produced by a population of relativistic particles accelerated in the shocks.}
   {Our goal is to investigate the occurrence of particle acceleration among massive stars in their pre-supernova evolution phases.
   }
   {We observed a subset of five WR stars in the radio domain using the upgraded Giant Metrewave Radio Telescope (uGMRT), located in India. The observations were carried out in bands 4 (550--950~MHz) and 5 (1050--1450~MHz) for all the targets. 
   }
   {We detected radio emission for only WR~110 in bands~4 and 5.
   Its thermal spectrum displays a consistent index of +0.74 down to uGMRT bands. The four other targets were not detected and we derived 3$\sigma$ upper limits. Our upper limits in Band~4 are the first provided for these targets below 1 GHz. None of the targets was identified as a synchrotron radio emitter in these radio bands.}
   {If some synchrotron emission is produced in these systems, the non-detection with uGMRT can be most likely attributed to strong free-free absorption (FFA). This is especially relevant for WR98a which is catalogued as a particle accelerator based on previous measurements at higher radio frequencies. The prominence of FFA constitutes a severe obstacle to identifying particle accelerators in the radio domain.}


\titlerunning{GMRT observations of massive stars}
\authorrunning{A. B. Blanco}
   \maketitle
%

\section{Introduction}

High-mass stars (M~$\geq$ 8~\(\textup{M}_\odot\)), that include O-type, and early B-type stars, as well as their evolved counterparts, the Wolf--Rayet (WR) stars, present strong and dense, radiatively driven stellar winds \citep[see e.g.][for a review]{Puls2008}.
These winds 
allow massive stars to continuously eject material from their outer layers, thus injecting chemically enriched matter and mechanical energy into the surrounding medium.
This ejected material is accelerated to terminal velocities ($\mathrm{v_{\infty}}$) of a few 1000~~$\mathrm{km\,s}^{-1}$
with mass-loss rates ($\dot{M}$) ranging typically between 10$^{-7}$ and 10$^{-5}$~$\mathrm{M_{\odot}\,yr}$$^{-1}$, depending on the spectral type and luminosity class. 
    
Furthermore, massive stars are mostly found to be part of binaries, or higher multiplicity systems. While the fraction of massive stars in binary systems has not been well determined yet, recent estimates point to a binary frequency 
of at least 90~$\%$ for O-type stars in young clusters \citep{Offner2023}.
In the case of a binary system made of high-mass stars, the powerful winds of both stars will collide somewhere in between the stars, giving rise to what we call a colliding-wind binary (CWB). Among the group of CWBs, there is a subcategory of binaries that are capable of accelerating particles up to relativistic velocities, and are called particle-accelerating
colliding-wind binaries (PACWBs, \citealt{PACWB2013, DeBecker2017}). 
To date, the catalogue of PACWBs includes about 50 objects \citep{PACWB2013}.\footnote{The updated version of the catalogue is available at \url{https://www.astro.uliege.be/~debecker/pacwb/}.}

The particle acceleration process that is active in PACWBs is believed to be diffusive shock acceleration (DSA; \citealt{Drury1983D}), occurring in the shocks produced in the colliding-wind region (CWR). This process can accelerate charged particles to relativistic velocities, including relativistic electrons at the origin of a series of non-thermal (NT) radiative processes.

Among these processes, it is worth emphasizing the importance of synchrotron emission. In the presence of a magnetic field, of likely stellar origin, relativistic electrons produce NT synchrotron radio emission, which is the main indicator of particle acceleration in PACWBs \citep[e.g.][]{PACWB2013}. Focusing on that spectral domain is thus especially relevant to investigating the occurrence of particle acceleration in massive star systems. In this framework, two specific motivations at the core of the present study deserve to be clarified.

First, despite significant advances in our understanding of PACWBs, a notable gap exists in our knowledge of the fraction of particle accelerators among massive binaries. 
Achieving a better view of the full population of PACWBs is especially relevant for understanding the contribution of these objects to the acceleration of Galactic cosmic rays \citep{DeBecker2017}. Second, since synchrotron radiation from massive stars is produced in the shocks taking place in the CWR, 
its detection in the radio domain points to a binary nature. As a result, binaries that are still unrevealed by classical multiplicity study methods (spectroscopic analysis, high angular resolution direct imaging...) may be evidenced by a particle acceleration signature. The connection between the NT radio emission from massive stars and their multiplicity is what we refer to as the synchrotron-binarity correlation. Keeping the aforementioned motivations in mind, we selected a sample of five WR stars for observation using the upgraded Giant Metrewave Radio Telescope (uGMRT). 

This paper is structured as follows. We present the sample of targets observed and give a brief description of them in Sect.\,\ref{Sample}. The data acquisition and reduction procedure are described in Sect.\,\ref{Obs}. In Sect.\,\ref{Res} we present the main results of the data analysis. A detailed discussion of the results and their implications is carried out in Sect.\,\ref{Disc}, while the main conclusions are summarized in Sect.\,\ref{Conc}, along with potential suggestions for future research in this area.

\section{Selected Wolf--Rayet stars}\label{Sample}

\subsection{Sample description}

We selected five WR~stars from the Galactic Wolf Rayet Catalogue\footnote{\url{https://pacrowther.staff.shef.ac.uk/WRcat/}} \citep{Rosslowe&Crowther2015}: WR~87, WR~93, WR~98a, WR~106, and WR~110. 
The motivation to select WR stars is two-fold. Firstly, WR stars are characterized by especially high values of the wind kinetic power (defined as the kinetic energy transferred per unit time in the wind outflow; $P_{\mathrm{kin}} = \frac{1}{2} \dot{M} \mathrm{v_{\infty}}^{2}$), typically of the order of (or even above) 10$^{37}$~$\mathrm{erg\,s}^{-1}$. This represents a significant reservoir of mechanical energy available to drive physical processes associated with their winds, in particular processes related to shock physics in CWBs such as particle acceleration and emission of synchrotron radiation. Secondly,  WR components are observed in nearly half of the identified PACWBs. This proportion is noteworthy, given that WR stars are expected to be remarkably less numerous than their O-type progenitors, considering the respective duration of their evolutionary phases. Additionally, the selected sample has shown potential indications of binarity, based on previous observational studies at different wavelengths. To mitigate the effects of brightness dilution associated with larger distances, we restricted our selection to stars within a maximum distance of 3~kpc. The selected sample of WR~stars is presented in Table~\ref{Targets}.

WR~87 and WR~110 are both classified as nitrogen-rich WN spectral types. The former has shown both absorption lines of unknown origin and weak emission lines (\citealt{CM89, Smith1996, VanDerHucht2001}). 
WR~110 shows a thermal X-ray emission spectrum harder than expected for a single star's wind, which could be explained by colliding wind shocks. In contrast, its radio emission was observed to be consistent with free-free (FF) emission from the ionized wind \citep{Skinner2002}. WR~110 has also been studied by \citet{Shara2022} in the search for a close and faint companion. However, they found no indication of any companion that would fit the criteria of their analysis. 

WR~93, WR~98a, and WR~106 belong to the carbon-rich WC spectral type. WR~93 is believed to be part of a binary system, with a possible late O-type companion revealed through spectroscopy studies showing double absorption lines, but without an orbital solution known to date (\citealt{Lortet1984, VanDerHucht2001}). 
WR~98a is an interesting target in the sample. It is confirmed as a PACWB \citep{PACWB2013} with evidence of NT radio emission (a variable flux, and a value for the spectral index $\alpha$ close to zero; \citealt{Monnier2002, Cappa2004}). This system is also observed to be associated with a dusty spiral pinwheel nebula revolving with a 1.4~yr periodicity \citep{Monnier1999}. The flux variation in the infrared (IR), believed to be due to the dust formation in the wind-wind interaction region, can be attributed to the orbital eccentricity of the system or to optical depth effects related to our viewing angle of the hot inner parts of the system \citep[see e.g.][]{ Williams1995, Monnier2002}. Hence, monitoring this system for NT radio emission is crucial to understanding the system in detail. WR~106 also belongs to the category of dust-makers, identified as such based on the detection of a significant IR excess, but not detected in X-rays \citep{CohenandVogel1978, deBecker2015}. While the non-detection of WR~106 in X-rays, plus the lack of Balmer absorption lines in its spectrum attributable to a hot companion \citep{Williams2000}, point to a single WC-type star scenario, the persistent production of dust is commonly explained in terms of the existence of enhanced density regions. These regions result from the collision of two winds, which would favour a binary system scenario. If WR~106 is not a CWB, a new explanation for the formation of its circumstellar dust should be found \citep{Williams2014}.

\subsection{Previous radio measurements}

All the targets studied here had been previously observed in the radio domain and other wavelengths, and thus benefit from past radio measurements. For some of them, we also found prior observations at 1.4~GHz, which is the frequency range compatible with the uGMRT Band~5. In the cases of 1.4~GHz observations, no emission has been detected either, with the previous works providing an upper limit.
Radio measurements of the targets available in the literature are given in Table~\ref{OtherMeasurements}. We only show values up to 15~GHz since the emission is primarily thermal at higher frequencies.
No measurements below 1 GHz have been reported for these targets. Hence, our data in Band~4 provide us with the first opportunity to measure the emission from these targets in this low-frequency range.

\begin{table*}[h]

\small\centering
\caption[]{Selected sample of WR~stars.}

\begin{tabular}{lcccccc}
\toprule

\label{Targets}

Source & $\mathrm{Spectral~Type}$ & $\dot{M}$  & $\mathrm{v_{\infty}}$ & $T_{\mathrm{eff}}$ & $D$ & $P_{\mathrm{kin}}$\\
 & & $(10^{-5} \mathrm{M}_{\odot} \mathrm{yr}^{-1})$ & $(\mathrm{km} \mathrm{s}^{-1})$ & (kK) & (kpc) & $(10^{37}\mathrm{erg}~\mathrm{s}^{-1})$ \\
\midrule

WR 87 & $\mathrm{WN7h~+~?}$ & 2.5$ ^{~\mathrm{a}}$  & $1400^{~\mathrm{a}}$  & $44.7^{~\mathrm{b}}$  & $3.0^{~\mathrm{c}}$ & $1.6$ \\


WR 93 & $\mathrm{WC7+O7-9?}$ & 2.0$ ^{~\mathrm{d}}$  & $2200^{~\mathrm{d}}$   & $75^{~\mathrm{d}}$   &  $1.76^{~\mathrm{e}}$ & $3.1$ \\


WR 98a & $\mathrm{WC8-9vd~+~?}$ & 1.6$^{~\mathrm{f}}$   & $1400^{~\mathrm{f}}$  & $60^{~\mathrm{d}}$ & $1.9^{~\mathrm{g}}$ & $1.0$ \\


WR 106 & $\mathrm{WC9d~+~?}$ & 1.6$^{~\mathrm{h}}$ & $1100^{~\mathrm{h}}$ & $45^{~\mathrm{h}}$ & $2.32^{~\mathrm{i}}$ & $0.6$ \\


WR 110 & $\mathrm{WN5-6b~+~?}$ & 1.6$^{~\mathrm{j}}$ & $2300^{~\mathrm{b}}$ & $70.8^{~\mathrm{b}}$ & $1.8^{~\mathrm{c}}$ &  $2.7$ \\

\bottomrule
\end{tabular}

    \begin{tablenotes}
      \small
      
      \item $\bf{References}$: 
      a)\cite{Hamann1995}, 
      b)\cite{Hamann2006}, 
      c)\cite{Crowther2023}, 
      d)\cite{Crowther2007}, 
      e)\cite{Rate&Crowther2020}, 
      f)\cite{PACWB2013}
      g)\cite{Monnier1999}
      h)\cite{Sander2019},
      i)\cite{VanDerHucht2001},
      j)\cite{Chene2011}.

      The spectral type was taken from the catalogue compiled by \cite{Rosslowe&Crowther2015} in all cases.
      The values for the kinetic power of the winds were derived in this work.
    \end{tablenotes}

\end{table*}


\begin{table*}[h]

\small\centering
\caption[]{Previous radio measurements.}

\label{OtherMeasurements}

\begin{tabular}{lcccc}
\toprule

Source & $S_{\nu}$(1.4 GHz) &  $S_{\nu}$(4.9 GHz)  & $S_{\nu}$(8.4 GHz) & $S_{\nu}$(15 GHz) \\
 & (mJy) & (mJy) & (mJy) & (mJy) \\
 
\midrule

WR 87 & < 1.26$^{~\mathrm{a}}$ & < 0.40$^{~\mathrm{b}}$  & < 0.24$^{~\mathrm{c}}$   &  \\


WR 93 &  & 0.9 $\pm$ 0.2$^{~\mathrm{b}}$ &  &  \\


WR 98a  & < 0.36$^{~\mathrm{d}}$  & 0.37 $\pm$ 0.07$^{~\mathrm{d}}$  & 0.47 $\pm$ 0.05$^{~\mathrm{c}}$    & 0.64 $\pm$ 0.11$^{~\mathrm{e}}$ \\
        &                         & < 0.39$^{~\mathrm{e}}$           & 0.55 $\pm$ 0.15$^{~\mathrm{d}}$    &  \\
        &                         &                                  & 0.59 $\pm$ 0.06$^{~\mathrm{e}}$    &  \\
        &                         &                                  & 0.62 $\pm$ 0.05$^{~\mathrm{d}}$    &  \\


WR 106 &  & &  < 0.17$^{~\mathrm{c}}$ & \\


WR 110 & & 1.17 $\pm$ 0.04$^{~\mathrm{f}}$ & 1.77$^{~\mathrm{f}}$ & \\

\bottomrule
\end{tabular}

    \begin{tablenotes}
      \small
      
      \item $\bf{References}$: 
      a)\cite{Chapman1999}, 
      b)\cite{Abbott1986},
      c)\cite{Cappa2004},
      d)\cite{Monnier2002},
      e)\cite{Montes2009},
      f)\cite{Skinner2002}.
   
          There are several values for the fluxes of WR~98a since it has been catalogued as a variable source.
    \end{tablenotes}

\end{table*}

\section{Observations and data reduction}\label{Obs}

The selected sample of WR~stars was observed with the uGMRT situated in Pune, India. This Y-shaped array consists of 30~parabolic dishes with a diameter of 45~m each. Each of the three arms of the array contains six antennas, while the remaining 12 dishes are distributed in a central square kilometre region. The antennas are fully steerable and span over an extension of 25~km, with 100~m being the minimum baseline length, which allows the array to probe small and large spatial scales of radio emission. It operates at frequencies ranging from 50~MHz to 1450~MHz.

Single-epoch observations of the five targets were carried out from June 4 to 9, 2023, during cycle 44. Continuum observations were obtained in Band~4 (550--950~MHz) and Band~5 (1050--1450~MHz) using the GMRT Wideband Backend (GWB) correlator set to have a bandwidth of 400~MHz covering 4096 channels. The primary flux calibrators observed were radio sources 3C48 and 3C286, while sources 1822--096, 1626--298, and 1911--201 served as phase calibrators for determining the amplitude and phase gains necessary for flux and phase calibration. 

All the steps of the data reduction process, namely flagging and removal of bad data and radio frequency interference (RFI) contaminated data, calibration, imaging of the target based on the calibrated visibility data, and self-calibration, were carried out using the \texttt{CAPTURE}\footnote{\url{http://github.com/ruta-k/CAPTURE-CASA6.git}} continuum imaging pipeline developed for uGMRT data \citep{Kale2021}. The pipeline makes use of tasks from the Common Astronomy Software Applications (\texttt{CASA}; \cite{McMullin2007}) software. For some targets, additional manual flagging had to be carried out. Six iterations of phase-only self-calibration were performed before achieving the final image, to obtain maps with better rms noise. The maps were subsequently primary beam corrected, using the task \texttt{ugmrtpb}.\footnote{\url{https://github.com/ruta-k/uGMRTprimarybeam-CASA6}} 
  
%


\section{Results}\label{Res}

The synthesized beam size in both bands is of the order of a few arcseconds. At a distance of a few kiloparsecs, this angular size corresponds to projected linear distances on the plane of the sky of thousands of astronomical units. Only binary systems with orbital periods of several thousand years would reach such dimensions, which are significantly longer than the orbital periods of all known PACWBs. Our targets, if detected, should thus be unresolved and appear as point sources. The radio maps were therefore inspected to reveal the presence of point sources at the expected position of our targets, above a detection threshold set based on the rms sensitivity of the maps. 

   \begin{figure*}
   \centering
    \includegraphics[width=0.9\hsize]{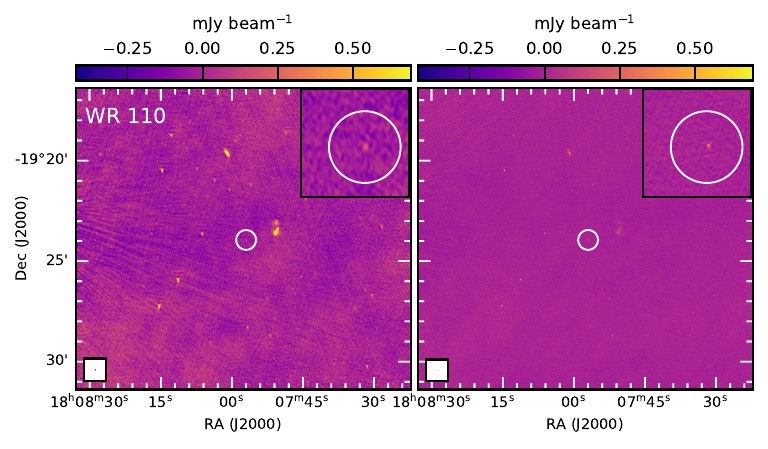}
      \caption{Field around WR~110 in Band~4 (left) and Band~5 (right). The location of WR~110 is given by the white circle. A zoomed-in view of the source is shown in the top right corner. 
              }
         \label{Images-WR110}
   \end{figure*}


\begin{table*}[h]
\small\centering
\caption[]{Parameters estimated from the radio maps}
\label{MapsDetails}
\begin{tabular}{cccccccc}
\hline
 \noalign{\smallskip}
 & \multicolumn{3}{c}{Band~4 (550-950~MHz)} & & \multicolumn{3}{c}{Band~5 (1050–1450~MHz)}\\
 \noalign{\smallskip}
\cline{2-4}\cline{6-8}
 \noalign{\smallskip}
Source & Angular resolution & rms & $\mathrm{S_{\nu}}$ & & Angular resolution & rms & $\mathrm{S_{\nu}}$ \\ 
  &  & (${\mu}$Jy $\mathrm{beam}^{-1}$) & (mJy) & &  & (${\mu}$Jy $\mathrm{beam}^{-1}$) & (mJy) \\
 \noalign{\smallskip}
\hline
 \noalign{\smallskip}
WR~87 & $7\farcs2 \times 3\farcs $ & 156 & $<$ 0.47 & & $3\farcs8 \times 1\farcs8$ & 29 & $<$ 0.09 \\ 
WR~93 & $8\farcs2 \times 3\farcs5$ & 478 & $<$ 1.43 & & $3\farcs5 \times 1\farcs9$ & 224 & $<$ 0.67 \\ 
WR~98a & $5\farcs9 \times 3\farcs4$ & 108 & $<$ 0.32 & & $4\farcs0 \times 1\farcs7$ & 34 & $<$ 0.10 \\ 
WR~106 & $5\farcs7 \times 3\farcs7$ & 106 & $<$ 0.32 & & $2\farcs9" \times 2\farcs0$ & 29 & $<$ 0.09 \\ 
WR~110 & $5\farcs6 \times 3\farcs6$ & 57 & 0.18 $\pm$ 0.08 & & $2\farcs6 \times 2\farcs1$ & 21 & 0.47 $\pm$ 0.05 \\ 
\hline
\end{tabular}
\end{table*}

Radio emission was detected only for WR~110 in both bands. The field maps are shown in Fig.~\ref{Images-WR110} with the location of the source encircled in white. No emission was detected for the other sources in both bands. The maps of these fields are presented in Fig.~\ref{RadioImages-NonDetections}. It is worth noting that we do not detect any radio emission from WR98a, which is reported as a PACWB from earlier work \citep{PACWB2013}. This is given a more specific discussion in Sect.\,\ref{wr98a}. In the cases of non-detection, we derive 3$\sigma$ upper limits to the flux density of the WR~stars from the rms levels at both frequencies. We use the CASA task \texttt{imfit} to estimate the flux density in both bands for WR~110. The faint detection in Band~4 is estimated to be just above the 3$\sigma$ level. The synthesized beam sizes, local rms values, and flux density values (and upper limits) of the clean Stokes I maps for all the targets in both bands are presented in Table~\ref{MapsDetails}.


\section{Discussion}\label{Disc}

\subsection{The radio emission from Wolf--Rayet stars}\label{radioWR}
The thermal and NT radio emission associated with massive stars, and particularly with WR~stars, in binary (or higher multiplicity) systems, has been extensively studied through the years \citep[e.g.][]{Dougherty2000, Montes2009, PACWB2013, DeBecker2019, Benaglia2020}. 
The radio emission from PACWBs is a combination of two sources of emission: (1) the FF thermal emission from the individual stellar winds, and (2), the NT synchrotron radiation produced by the population of relativistic electrons accelerated in the CWR. The thermal emission is roughly steady, though in a binary system some enhancement is expected of the thermal emission, due to the wind compression in the CWR, especially for close binaries and/or systems with similar mass-loss rates \citep{Stevens1995}. Variability of the thermal emission could in principle be observed for some binaries, with the variation associated with changes in the separation of the stars or orientation effects that result in a differential absorption through the circumstellar material \citep{Pittard2010}. The most noticeable example of variable thermal emission from a massive binary is $\eta$\,Car, where the variability is explained by an orbital, phase-dependent alteration of the ionization of the wind material as it is exposed to the intense radiation field from the bright companion \citep{Kashi2007}. For its part, NT emission produced in the CWR is intrinsically variable in eccentric orbits as a function of the orbital phase. It reaches its peak at periastron, due to a stronger local magnetic field acting at the CWR as a consequence of the smaller separation between the stars. In both cases, the emission spectra in radio follow a power-law dependence on the frequency and can be described by the expression $S_{\nu} \propto \nu^{\alpha}$.
The power-law exponent, $\alpha$, is known as the spectral index and is typically below -0.1 for synchrotron radiation \citep[e.g.][]{Kellerman1964}, and canonically 0.6 \citep{WB1975, PF1975} for thermal FF radio emission assuming a spherically symmetric and homogeneous, single-star wind. This implies that the synchrotron radiation is stronger and dominates the radio spectrum at lower frequencies. At the same time, the thermal contribution from the winds increases with the frequency, and thus this emission dominates at higher frequencies. Consequently, a negative spectral index measured in the radio spectrum of a source and/or a bright emission in the low-frequency range indicates the emission's synchrotron nature. 

However, synchrotron radio emission produced in the CWR is likely to be attenuated 
by free-free absorption (FFA). This absorption process is stronger at lower frequencies and depends on the column of ionized stellar wind along the line of sight. It is thus more pronounced in short-period binary systems, at times leading to a complete suppression of the synchrotron component, especially when the absorbing stellar winds are dense \citep[e.g.][]{delPalacio2016, DeBecker2019}. In this context, WR winds are the most efficient absorbers. Thus, WR winds, with their high kinetic power, can feed NT processes efficiently and are, at the same time, the most likely to absorb the synchrotron emission through FFA if the stellar separation is small enough. The FFA process is sensitive to orientation effects and the eccentricity of the orbit, as it depends on the column density of absorbing material along the line of sight. Consequently, FFA undergoes orbital modulation and intensifies as the system approaches periastron and the stellar separation decreases. This effect competes with the enhancement of synchrotron emission, also occurring at periastron.

Another phenomenon that can suppress the synchrotron radiation produced in a thermal plasma at low frequencies is the Razin effect (also known as the Razin--Tsytovich effect). This effect generates an exponential, sharp cutoff in the synchrotron radio spectrum, typically at frequencies lower than the Razin-Tsytovitch cutoff frequency. According to \citet{Pacholczyk1970}, this frequency (in Hz) can be expressed as $\nu_{\mathrm{R}} = 20\,n_{\mathrm{e}}/B$, where $n_{\mathrm{e}}$ is the number density of electrons in the plasma and $B$ is the magnetic field intensity (both quantities expressed in cgs units). An in-depth exploration of the parameter space of every specific system is required to estimate which of these two processes (FFA or Razin effect) dominates the turnover. This was addressed for instance in the detailed modelling approach presented by \citet{Dougherty2003}. Given the stronger sensitivity of FFA on the density, which is significantly high in WR systems (the absorption coefficient scales with the density squared) compared to the Razin effect (cutoff frequency proportional to the density), we focus here on FFA only. Addressing more deeply the question of the nature of the dominant turnover process would require a much better knowledge of the parameters characterizing our targets.


\subsection{The multiplicity of WR stars}\label{multWR}

Most WR binary stars are observed to have an OB star companion, some of them were found to have compact companions \citep[e.g.][]{Crowther2010a}, and only a few binary systems made of two WR~stars are known so far, for example Apep, discovered by \citet{Callingham2020}.
The multiplicity fraction among WR stars was thought to be approximately 0.4, as reported by \citet{VanDerHucht2001}. This value encompassed both spectroscopic and visual WR binaries but was strongly biased towards short-period systems. However, \citet{Dsilva2020} carried out a bias-corrected study of the multiplicity properties of Galactic WC-type WR stars 
obtaining an intrinsic multiplicity fraction for the Galactic WC population of at least 0.72. 
In a later study, they analysed the early-type, nitrogen-rich WR population \citep[WNE,][]{Dsilva2022} and the late-type, nitrogen-rich WR population \citep[WNL,][]{Dsilva2023} of the Galaxy. 
The combined study of both types of Galactic WN populations yields an intrinsic binary fraction for WN stars of 0.52, significantly lower than that obtained for WC stars.  

Generally, detecting potential companions of WR stars can be a challenging task due to the inherent characteristics of these stars \citep[e.g.][]{Shenar2022, Shara2022}. 
They emit intense radiation across a wide spectrum and this high intrinsic brightness makes them likely to overwhelm, by contrast, a putative, comparatively fainter emission from their companions in binary systems. 
The difficulty in detecting WR companions renders indirect methods, for example the synchrotron-binarity correlation, more valuable. Since FFA is more prominent in short-period binary systems, where stellar separation is small, detecting synchrotron radio emission from WR systems would reveal mostly systems with sufficiently long periods, typically spanning months, and often extending to years or even decades. This is particularly important considering that most studies and techniques focus on short-period systems (days, weeks, months).

The fact that synchrotron radiation from massive stars is thought to be produced almost exclusively in the context of a binary system \citep{Dougherty2000, PACWB2013} allows us to investigate the question of their multiplicity in a region of the orbital parameter space that is not so well explored and contributes to a better census of long-period binaries among WR stars.


\subsection{Thermal emission}\label{sectT}

\subsubsection{Detection of WR\,110}
\begin{figure}
   \centering
    \includegraphics[width=1.0\hsize]{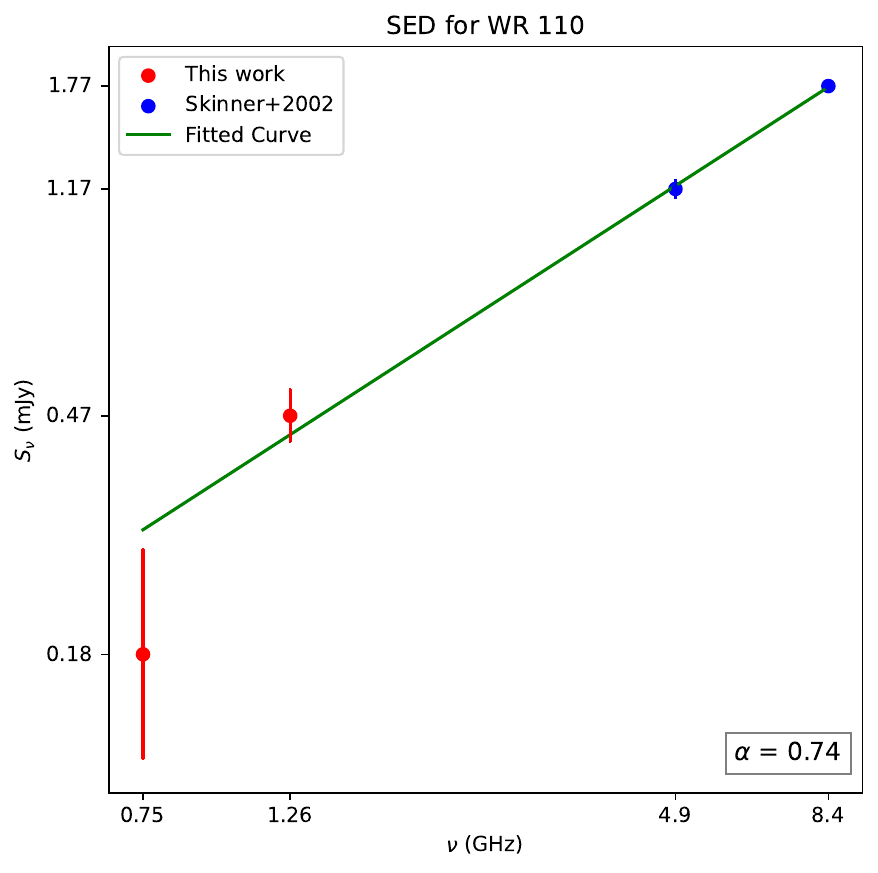}
      \caption{Spectral energy distribution for WR~110 based on our observations and the ones from the literature. The fitted curve gives a spectral index $\alpha$ = 0.74 $\pm$ 0.01.
              }
         \label{SED-WR110}
   \end{figure}

The detection of low-frequency radio emission in WR\,110 is worth comparing with other measurements at higher frequencies (see Table\,\ref{OtherMeasurements}). In particular, we expect higher frequency radio emission to be dominated by the thermal emission from the dense WR wind. The radio spectral energy distribution (SED) including our uGMRT measurements is shown in Fig.\,\ref{SED-WR110}. The radio emission we detect in Band~5 is fully consistent with the trend displayed by previous flux density measurements at 4.9 and 8.4\,GHz. This suggests that a unique radio emission process is at work across the SED. We used our measurements at bands 4 and 5 together with measurements at higher frequencies to estimate the spectral index $\alpha$ in this frequency region and obtained a value of 0.74 $\pm$ 0.01. The positive slope of the spectrum points to thermal emission from the wind, with a spectral index greater than the canonical value of 0.6 predicted by \citet{WB1975}. However, this is in agreement with other thermal indices determined for WR stars, and results from the influence of strong clumping in WR winds, as demonstrated by \citet{Nugis1998}. We note that the flux measurement in Band~4 is lower than the linear trend displayed by higher frequency values. Given that the significance level in the Band~4 measurement is low, it would be purely speculative to provide a valid physical interpretation of this low-frequency deviation. We do not anticipate a change in slope at low frequencies, despite the limited investigation of WR stars in this range. A notable example is WR~11, which exhibits a consistent thermal spectrum down to 150\,MHz  \citep{Benaglia2019}.

\subsubsection{Non-detections}
For WR\,87, WR\,93, and WR\,106, the lack of detection at 1.4\,GHz and higher frequencies by previous studies (Table\,\ref{OtherMeasurements}) prevents us from estimating a spectral index. For WR\,98a, its already established PACWB status does not allow us to determine a thermal index as we did for WR\,110. As a consequence, it is not possible to investigate whether these WR stars also deviate from the canonical thermal emission law derived by \citet{WB1975}. 

Even though we have good reasons to consider that our targets deviate from the canonical behaviour, as explained by \citet{Nugis1998}, we predict the order of magnitude of the expected thermal flux density ($S_{\nu}^{\mathrm{ff}}$) in the wind of the stars using a slightly modified version of the equation given by \citet{WB1975}:

\begin{equation} 
\label{eqn:thermalFlux}
\begin{aligned}
{
\left[\frac{S_{\nu}^{\mathrm{ff}}}{\mathrm{Jy}}\right]=  23.2\left(\left[\frac{\dot{M}}{\mathrm{M}_{\odot} \mathrm{yr}^{-1}}\right]\left[\frac{\mathrm{km}~\mathrm{s}^{-1}}{\mathrm{v_{\infty}}}\right] \frac{1}{\sqrt{f_{\mathrm{w}}} \mu}\right)^{4 / 3}  \times\left(\gamma g_{\mathrm{ff}} Z^{2} {\nu}\left[\frac{\mathrm{kpc}}{D}\right]^3\right)^{2/3}
}
\end{aligned}
\end{equation}

\noindent  
The values used for each target are listed in Table~\ref{Targets}.
The stellar wind clumping is considered in the volume filling factor $f_{\mathrm{w}}$; the parameters $\mu$, Z, and $\gamma$ are the mean molecular weight, the rms ionic charge, and the mean number of electrons per ion in the wind, respectively; and $D$ is the distance to the source. For the calculations, we assumed the following set of values: $f_{\mathrm{w}}=4$, $\mu=2$, $\mathrm{Z}=1$, and $\gamma=1$. The free--free Gaunt factor, $g_{\mathrm{ff}}$, was approximated, as done in \cite{Leitherer91}, 

\begin{equation}
g_{\mathrm{ff}} = 9.77\left(1+0.13 \log \frac{T_\mathrm{e}^{3 / 2}}{Z \nu}\right) 
\end{equation}

\noindent where the electron temperature of the wind was estimated as $T_\mathrm{e} \approx 0.3 T_{\mathrm{eff}}$ for the outer layers of the wind where the radio emission is detectable \citep{Drew1990}, and the values for the effective temperature are quoted in Table~\ref{Targets}. The predicted flux density values for each non-detected target in both bands are listed in Table~\ref{PredictedFF}.


\begin{table}[h]

\small\centering
\caption[]{Predicted values for the thermal flux density due to FF emission of the winds.}

\begin{tabular}{lcc}
\toprule

\label{PredictedFF}

Source & $\mathrm{S_{\nu}^{ff}}$ (Band~4) & $\mathrm{S_{\nu}^{ff}}$ (Band~5) \\
  & (mJy) & (mJy) \\

\midrule

WR 87 & 0.40 & 0.54  \\


WR 93 & 0.42 &  0.58 \\


WR 98a & 0.48 &  0.66 \\


WR 106 & 0.17 & 0.23  \\

\bottomrule
\end{tabular}

\end{table}


Comparing these theoretically estimated values with the 3$\sigma$ upper limits presented earlier in Table~\ref{MapsDetails}, we find the following: 

\begin{itemize}
    \item For the targets WR~87, WR~93, and WR~106 in Band~4, and WR~93 in Band~5, we obtain values for the estimated fluxes that fall below the measured 3$\sigma$ upper limit. In this sense, they are consistent with pure thermal emitters, and their expected emission is in agreement with a non-detection on our side. 
    \item For WR~87 and WR~106 in Band~5, and WR~98a in both bands, the model predicts flux density values that are somewhat higher than our upper limits. However, these differences are not substantial when considering the inherent uncertainties of this approach. Given the uncertainties in the mass-loss rates and the influence of clumping, which is not properly accounted for in this method, some deviations are anticipated. Specifically, it is important to highlight that the actual thermal emission from these WR winds is likely steeper than predicted by the canonical law assumed in Eq.\,\ref{eqn:thermalFlux}, leading to a lower flux at the low frequencies probed by our uGMRT data. We can thus conclude that our upper limits are not at odds with the expected behaviour of WR thermal winds.
\end{itemize}


\subsection{Non-detection of NT emission} \label{ntsection}

\subsubsection{Interpretation framework}
Although detecting synchrotron radio emission in our study would have provided strong evidence of particle acceleration processes near the star (indicating that the star is part of a multiple system), the absence of 
emission does not rule out this possibility either. A similar situation was encountered recently by \cite{Saha2023}, and we adopt their approach to analyse the results.

The non-detection of radio emission in the lower frequency bands of the uGMRT, despite their relevance for probing the NT synchrotron emission, could be due to a few different reasons worth discussing.


The stars in our sample might not be part of a binary system with a massive companion. If this were the case, there would not be wind shocks that provide the required conditions for accelerating particles, and thus no synchrotron emission. The possibility that the targets are isolated WR~stars and not part of a CWB cannot be discarded. In this case, we would only expect the thermal FF emission from the ionized wind of a single star, which may not be strong enough to be detected in the low-frequency observational bands used here.


If we assume that they are part of massive binary systems, one plausible justification for the non-detection of synchrotron radiation could be due to a binary configuration characterized by a small separation, in a short-period system. If this were the case, the CWR would be deeply embedded within the radio-opaque stellar winds and the synchrotron emission would result significantly reduced due to FFA.

      
Considering a binary system, but with a significantly longer period than considered above, the lack of detection of synchrotron emission could arise in some parts of a highly eccentric orbit. It is acknowledged that the emission from the CWR depends on the stellar separation and that the synchrotron emission peaks towards the periastron, where the stellar separation is the smallest. In the case of a very long-period and highly eccentric orbit, two different configurations could lead to a lack of detection. In one case, if the stars of the system are close to apastron at the time of our observations, the large distance between them would cause the synchrotron emission to drop to its minimum. There are two reasons for this. The first is a drop in the local magnetic field (of stellar origin) in the CWR since it decreases as the separation between the stars increases \citep{Usov&Melrose1992} and the second is a drop in the injection rate of particles into the DSA process due to a lower density of the colliding flows in the case of a larger separation \citep{deBecker2024}. This could lead the synchrotron flux to fall below the detection threshold. Alternatively, close to periastron, the stellar separation can be small enough to substantially increase the impact of FFA, up to a level that may suppress any synchrotron radiation produced in the deeply embedded CWR, as described in the previous case. In this sense, the variable nature of the radio NT emission produced in the wind--wind interaction region draws attention to the temporal aspect of our observations, and it is worth noting that observations of the same target taken at a different epoch could lead to a very different result in terms of detection of NT emission.

As FFA could potentially explain the lack of synchrotron radio emission (provided it is produced in the system), it is worth looking at the constraints this provides in terms of the size of the putative binary system. We adopted the formalism proposed by \citet{WB1975} making use of the wind parameters provided in Table\,\ref{Targets} to estimate the radius equivalent to an optical depth ($\tau$) equal to one for the WR winds. We adopted the same assumption for the temperature as in Sect.~\ref{sectT}. Given the lack of information on putative companions, and the much weaker contribution to FFA coming from OB-type winds, we restrict the discussion to the WR winds. 

\begin{figure}
   \centering
    \includegraphics[width=1.0\hsize]{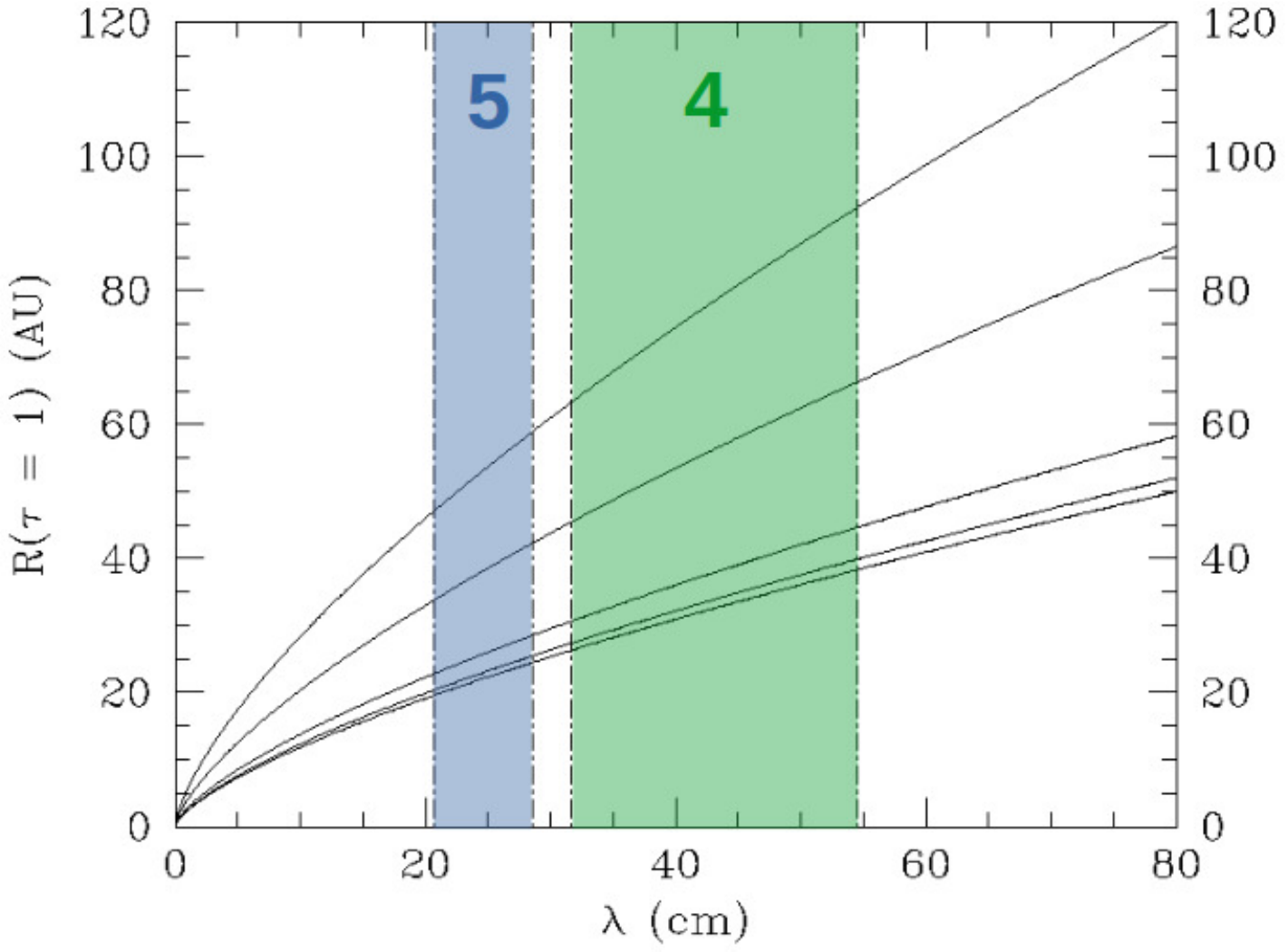}
      \caption{Extension of the radio photosphere as a function of wavelength for the five targets. The uGMRT bands are highlighted in green (Band\,4) and blue (Band\,5). From top to bottom: WR 87, WR 106, WR 98a, WR 110, and WR 93.
              }
         \label{radiophot}
   \end{figure}

Figure~\ref{radiophot} shows the effective radii of the stellar wind photosphere as a function of the wavelength; uGMRT Band~5 gives a tighter constraint. The most conservative interpretation of these curves is to consider that a separation between the WR and the synchrotron emitting region shorter than $R(\tau = 1)$ should lead to a significant level of FFA. Separations of a few tens of AU at a few kiloparsecs in principle convert to orbital periods of several years. However, various aspects have to be considered here. First, the synchrotron emission region is quite extended, opening the possibility for a fraction of the NT emission to escape the WR wind. The system's orientation is also expected to play a role: a WR wind located at the background of the NT emission region may allow some synchrotron emission to escape towards the observer. Finally, wind clumpiness is also likely to make the wind less opaque. All these factors may allow the synchrotron emission component to be detected even for smaller separations than plotted in Fig.\,\ref{radiophot}.
 
\subsubsection{The case of WR~98a}\label{wr98a}

\begin{figure}
   \centering
    \includegraphics[width=1.0\hsize]{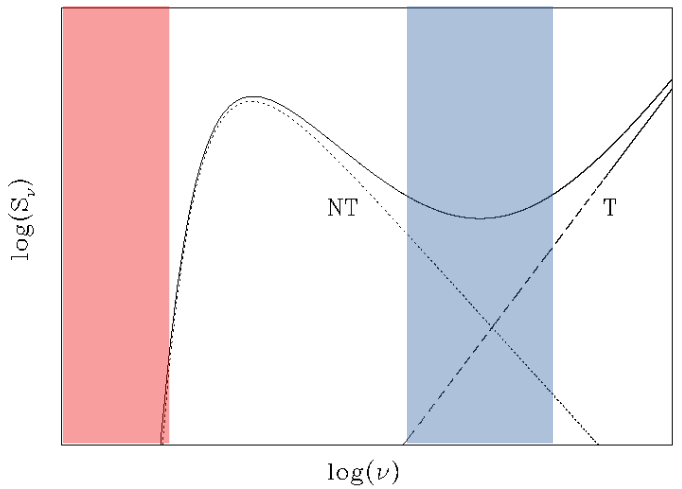}
      \caption{Schematic view of the radio composite spectrum (solid curve) of a PACWB presenting both a thermal component (T, dashed line) and a turned-over synchrotron component (NT, dotted curve). The blue highlighted part covers a spectral region with an almost flat spectrum, while the red part covers the region where NT emission is suppressed.
              }
         \label{wr98acolor}
   \end{figure}

As a bona fide PACWB, the non-detection of radio emission from WR~98a deserves further discussion. The synchrotron emitter status was established based on radio measurements at frequencies of several GHz \citep{Montes2009}. In particular, flux densities at 8.4 and 15 GHz (Table\,\ref{OtherMeasurements}) indicate a flat spectrum (spectral index close to 0.0). This typically is evidence of a composite spectrum, where both thermal and NT emission components contribute to the radio measurement. This can be better appreciated in the generic spectrum illustrated in Fig.\,\ref{wr98acolor}, which enables a qualitative discussion. 
The part of the composite spectrum that appears flat is highlighted in blue. The expectation for a brighter radio synchrotron emission at low frequencies (as discussed in Sect.~\ref{radioWR}) is valid when the optically thin unabsorbed part of the synchrotron spectrum is considered. However, turnover processes significantly suppress the radio emission in the lower part of the spectrum. The severe drop in the composite spectrum illustrates this. Our uGMRT measurements are possibly coincident with the spectral zone shown in red in Fig.\,\ref{wr98acolor}. In the case of WR~98a, FFA is most likely the turnover process at work, given the density and composition of the WC wind (see also Fig.\,\ref{radiophot}). According to \citet{Monnier1999}, this may indicate an orbital period of 1.4\,yr, which is short enough to make the synchrotron emission region severely affected by FFA over the majority of the orbit.
\citet{Hendrix2016} estimated the orbital separation of the system to be $\sim$ 4~AU, based on the orbital period and the stellar masses. Comparing this result to our estimation in Fig.~\ref{radiophot}, we can expect WR~98a to be completely embedded in the radio photosphere of the WR~star, which should be at least 20~AU.
In the scenario proposed here to reconcile our non-detection with previous results, the FFA turnover frequency should be well above 1.4~GHz (frequency of the Band~5), but below frequencies closer to 8.4 GHz.

Finally, we note that the NT component's variability must be considered (see Sect.\,\ref{radioWR}). The position of the FFA turnover is in particular very likely to change as a function of the orbital phase. This is certainly the reason why measurements at 4.9~GHz at some epochs point to lower flux density than those measured at higher frequencies (Table\,\ref{OtherMeasurements}). This probably indicates that, at these epochs, the turnover occurred at a rather high frequency, potentially fully suppressing the synchrotron components and revealing only the thermal emission from the WR wind (dashed line in Fig.\,\ref{wr98acolor}).
   

\section{Conclusions}  \label{Conc}

We conducted a radio continuum analysis of five WR~stars (WR~87, WR~93, WR~98a, WR~106, and WR~110) that had previously displayed signs of being part of a multiple system. In the attempt to improve our comprehension of their possible role as particle accelerators, we performed low-frequency (735 and 1260~MHz) uGMRT observations in June 2023.  

No radio emission was detected from WR~87, WR~93, WR~98a, and WR~106. For these cases, we give 3$\sigma$ upper limits to the radio flux densities in both bands of observation. We provide the first-ever observations in Band~4, a frequency range unexplored for these targets. For WR~93, WR~106, and WR~110, we offer the first observations in Band~5. 

For WR~110, we detect Band~4 and Band~5 emissions consistent with pure thermal emission. Making use of our new measurements and of previously published flux density values at 4.9 and 8.4\,GHz, we derived a thermal spectral index of 0.74\,$\pm$\,0.01. This value, steeper than the usual canonical value, is compliant with expectations for dense and highly clumped WR winds. If WR~110 were indeed a binary system, and based on Fig.~\ref{radiophot}, we can see that its orbital separation should be greater than $\sim$ 20 or even 30~AU, depending on the observing frequency, for us to detect some NT emission.

We found no observational evidence of synchrotron radiation, and therefore no indication of binarity, in any of the targets. Our uGMRT measurements at low-frequency bands are likely significantly affected by FFA, which would heavily suppress any potential synchrotron emission at these frequencies. 
This is particularly notable in the case of WR~98a; although it shows a composite spectrum (made of thermal and NT contributions) at higher frequencies, it was undetected in our data. This underscores the prominent role of FFA on observational biases affecting the census of PACWBs among CWBs. 
Detecting synchrotron emission at lower frequencies is thus highly dependent on the position of the FFA turnover. Therefore, we note the importance of including measurements at a few GHz as well, to investigate the nature of the radio emission from massive stars and to improve the census of particle accelerators among them. However, we note that lower frequency measurements remain necessary to characterize the turnover, especially in binary systems with sufficiently long periods that push the FFA turnover frequency below 1 GHz. These factors must be considered when addressing the observation biases that impact our census of particle accelerators among massive stars.


\begin{acknowledgements}
We want to express our gratitude to the referee for a constructive and fair report that helped us improve the paper. This research is part of the PANTERA-Stars collaboration, an initiative aimed at fostering research activities on the topic of particle acceleration associated with stellar sources.\footnote{\url{https://www.astro.uliege.be/~debecker/pantera}} This research has made use of NASA's Astrophysics Data System Bibliographic Services. This publication benefits from the support of the French Community of Belgium in the context of the FRIA Doctoral Grant awarded to A. B. Blanco.
\end{acknowledgements}

\bibliographystyle{aa}
\bibliography{biblio}

\begin{appendix}

\section{Radio maps of the uGMRT fields}

\begin{figure}
   \begin{center}
   \begin{minipage}{10cm}
   \includegraphics[width=1.0\hsize]{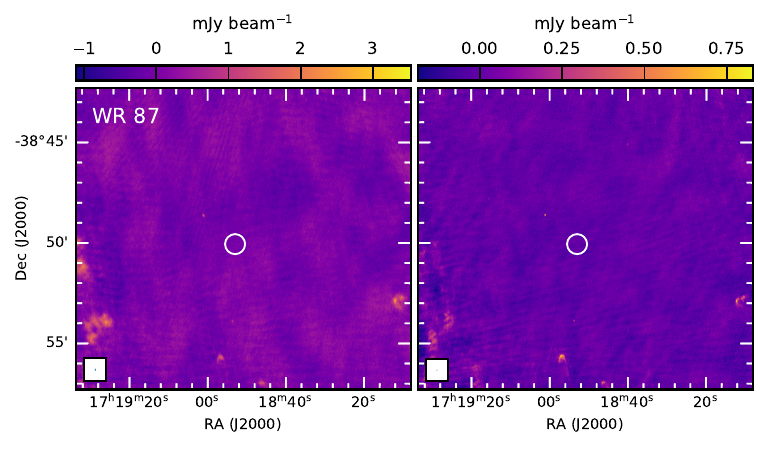}
   \end{minipage}\hfill
   \begin{minipage}{10cm}
   \includegraphics[width=1.0\hsize]{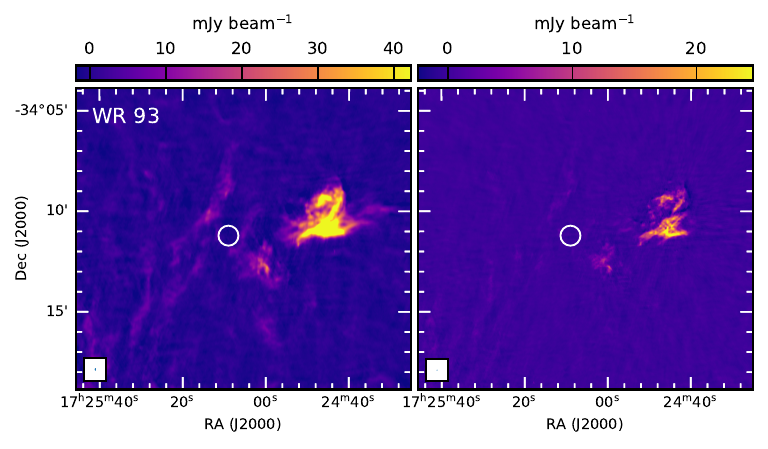}
   \end{minipage}
   \begin{minipage}{10cm}
   \includegraphics[width=1.0\hsize]{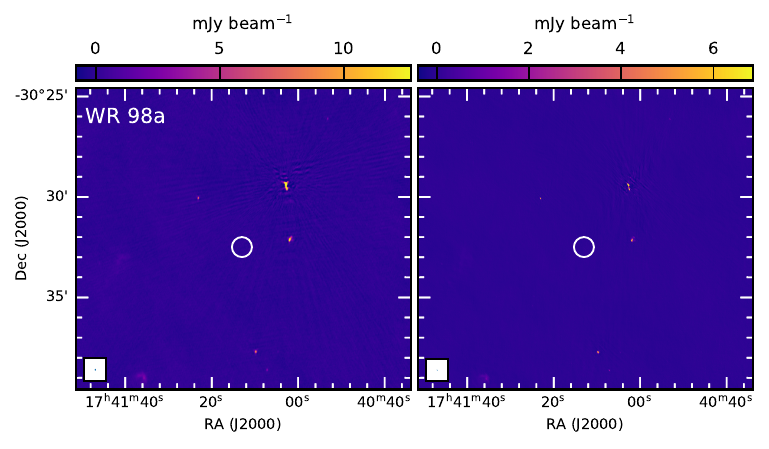}
   \end{minipage}\hfill
   \begin{minipage}{10cm}
   \includegraphics[width=1.0\hsize]{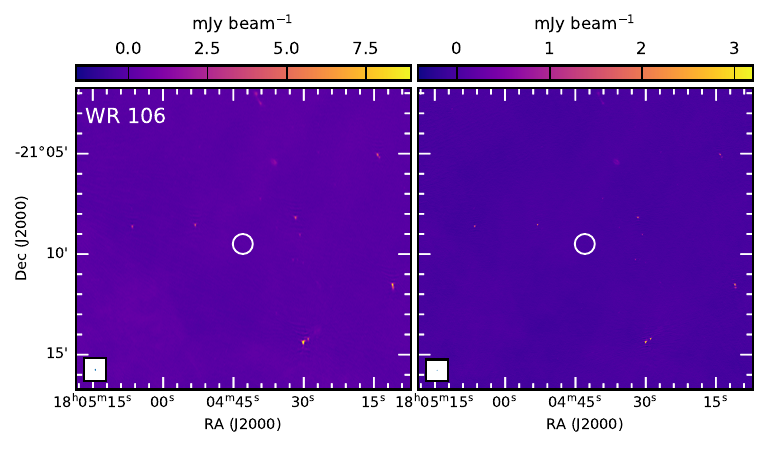}
   \end{minipage}
      \caption{uGMRT fields around WR~87, WR~93, WR~98a, and WR~106 observed in Band~4 (left column) and Band~5 (right column).}
         \label{RadioImages-NonDetections}
   \end{center}
   \end{figure}

Radio images without detection of the targets in Band~4 (left column) and Band~5 (right column) are shown in Fig.\,\ref{RadioImages-NonDetections}. In all the cases, the expected position of our target is indicated by a white circle. The synthesized beam is shown in the bottom left corner.

\end{appendix}

\end{document}